\pdfoutput=1

\documentclass[11pt]{article}

\usepackage[]{acl}

\usepackage{times}
\usepackage{latexsym}
\usepackage{hyperref}
\usepackage{multirow}

\usepackage[T1]{fontenc}

\usepackage[utf8]{inputenc}

\usepackage{graphicx}
\usepackage{graphics}
\usepackage{caption}
\usepackage{subcaption}
\usepackage{pdfpages}
\usepackage{float}
\usepackage{hyperref}

\usepackage{xcolor}
\usepackage[linesnumbered,ruled,vlined]{algorithm2e}
\title{Algorithm template}

\SetCommentSty{mycommfont}
\SetKwInput{KwInput}{Input}                
\SetKwInput{KwOutput}{Output}              

\usepackage{microtype}

\title{VendorLink: An NLP approach for Identifying \& Linking Vendor Migrants \& Potential Aliases on Darknet Markets}

\author{Vageesh Saxena \\
  Law \& Tech Lab \\
  Maastricht University \\
  v.saxena@maastrichtuniversity.nl \\\And
  Nils Rethmeier \\
  Speech and Language Technology Lab \\
  DFKI, Berlin  \\
  nils.rethmeier@dfki.de \\ \AND
  Gijs Van Dijck \\
  Law \& Tech Lab \\
  Maastricht University \\
  gijs.vandijck@maastrichtuniversity.nl \\
  \And
 Gerasimos Spanakis \\
  Law \& Tech Lab \\
  Maastricht University \\
  jerry.spanakis@maastrichtuniversity.nl \\
  }

\begin{document}
\maketitle
\begin{abstract}
The anonymity on the Darknet allows vendors to stay undetected by using multiple vendor aliases or frequently migrating between markets. Consequently, illegal markets and their connections are challenging to uncover on the Darknet. To identify relationships between illegal markets and their vendors, we propose VendorLink, an NLP-based approach that examines writing patterns to verify, identify, and link unique vendor accounts across text advertisements (ads) on seven public Darknet markets. In contrast to existing literature, VendorLink utilizes the strength of supervised pre-training to perform closed-set vendor verification, open-set vendor identification, and low-resource market adaption tasks. Through VendorLink, we uncover (i) 15 migrants and 71 potential aliases in the Alphabay-Dreams-Silk dataset, (ii) 17 migrants and 3 potential aliases in the Valhalla-Berlusconi dataset, and (iii) 75 migrants and 10 potential aliases in the Traderoute-Agora dataset. Altogether, our approach can help Law Enforcement Agencies (LEA) make more informed decisions by verifying and identifying migrating vendors and their potential aliases on existing and Low-Resource (LR) emerging Darknet markets. \footnote{Our code implementation is publicly available at \href{https://github.com/maastrichtlawtech/VendorLink.git}{https://github.com/maastrichtlawtech/VendorLink.git}
}
\end{abstract}

\section{Introduction}
\label{sec:introduction}

Conventional search engines index surface-web websites that only constitute 4\% of the entire internet \citep{dark-web-stats}. The remaining comprises 90\% Deep Web (not indexed) and 6\% Darknet, which uses advanced anonymity enhancing protocols \citep{dark-web-stats}. While the former serves legitimate purposes requiring anonymity, the latter is also used for illegal activities such as financial fraud \citep{financial-fraud}, child exploitation \citep{inbook}, and trading of illicit weapons \citep{10.2307/26297596, RR-2091-PACCS}, prohibited drugs, and chemicals \citep{RB-9925-WODC}.

Given the Darknet's scope, size, and anonymity, it is difficult for LEA to uncover connections between illegal marketplaces \citep{combating-crime}. While manual detection of such connections is a time-consuming and resource-extensive process, the recent success of online scrapers \citep{10.5555/1814490.1814502,hayes2018framework} and monitoring systems \citep{CyCon19,9029072} has enabled researchers and LEA to analyze \citep{10.2307/26783825,darStats,RR-2704-NIJ,doi:10.1177/0022018320952557} and automatically identify \citep{al-nabki-etal-2017-classifying,10.1145/3097983.3098193,eindhoven,10.1145/3322645.3322691} Darknet contents. This research proposes a vendor verification and identification approach to help LEA make better decisions by linking vendors, offloading manual labor, and generating similarity-based analyses. In contrast to the existing Darknet literature \citep{DBLP:journals/corr/HeZRS15, Ekambaranathan2018UsingST,10.1145/3292500.3330763, 10.1145/3366423.3380263, https://doi.org/10.48550/arxiv.2207.03477}, VendorLink, as illustrated in Figure \ref{fig:experiments}, emphasizes the following contributions to the problem of verifying and identifying vendors on Darknet markets: 

\begin{figure*}
    \centering
    \includegraphics[height=6.5cm, width=\linewidth]{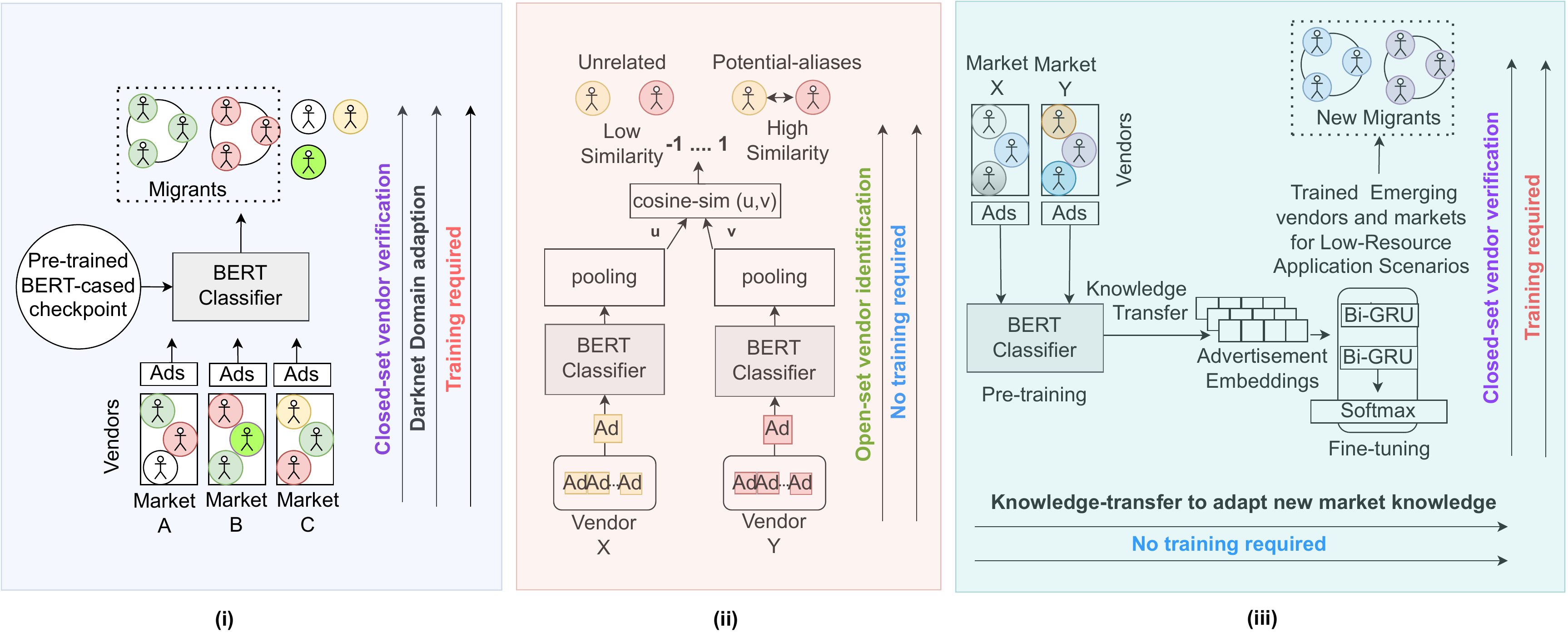}
    \caption{\label{fig:experiments} \textbf{(i)} \textbf{Closed-Set Vendor Verification Task:} A supervised pre-training task that performs classification using a BERT-cased classifier in a closed-set environment to verify unique vendor migrants across existing markets, \textbf{(ii)} \textbf{Open-Set Vendor Identification Task:} A text-similarity task in an open-set environment that utilizes style representations from the established BERT-cased classifier to verify known vendors and identify potential-aliases, \textbf{(iii)} \textbf{Low-Resource Market Adaptation Task:} A knowledge-transfer task in a closed-set environment to adapt new market knowledge and verify migrants across Low-Resource (LR) emerging markets.}
\end{figure*}

\paragraph{(i) Closed-Set Vendor Verification Task:} Due to limited resources, LEA prioritizes investigating Darknet vendors based on the size and nature of their trade. Thus, Darknet vendors often distribute their business across multiple markets to stay undetected. Likewise, some vendors relocate and resume their business in other markets after a market seizes \citep{9583681}. We refer to these migrating vendors as \textit{migrants} for brevity. Unfortunately, this movement prevents LEA from correctly estimating the size of a vendor's operations. To aid LEA, we perform supervised pre-training by conducting multiclass classification in a closed-set environment \cite {https://doi.org/10.48550/arxiv.2103.15086} to analyze different writing styles in text ads and classify vendor migrants to unique vendor accounts across three Darknet markets. Moreover, researchers have observed a significant difference in language structure between Darknet and Surface net websites \citep{choshen-etal-2019-language, https://doi.org/10.48550/arxiv.2204.06885}. Since most contextualized models are trained on surface web data, the supervised pre-training step allows our model to adapt to the Darknet market domain knowledge.

\paragraph{(ii) Open-set Vendor Identification Task:} Darknet vendors often create aliases and work in groups to distribute their products across multiple markets, allowing them to expand their business without being detected by LEA. Moreover, given the scope and anonymity of the Darknet, manually linking these profiles is infeasible. Hundreds of new markets and vendors emerge daily on the Darknet. While the existing literature has established impressive performance on the vendor verification task, any trained classifier will fail during inference to encounter unknown vendors from emerging markets in real-to-close-world scenarios. Therefore, in this research, we use the style representations from the pre-trained classifier to compute the cosine similarity between the text ads to verify existing vendors and identify potential aliases and unknown vendors in an open-set environment \cite {https://doi.org/10.48550/arxiv.2103.15086}.  

\paragraph{(iii) Low-Resource Market Adaptation task:} While research has demonstrated impressive performance for the Darknet's vendor verification task \citep{10.1145/3366423.3380263, https://doi.org/10.48550/arxiv.2207.03477}, high computational and storage requirements pose a significant challenge to LEA. Furthermore, with the exponential growth of Darknet markets and vendors with new content every year, there is a dire need for systems that can verify existing vendors from a known database and simultaneously adapt to new market knowledge from emerging vendors and markets. After all, not all LEA have the resources to train computationally expensive models from scratch. Therefore, this experiment investigates our classifier's capability to benefit transfer learning in a low-resource setting \citep{ruder-etal-2019-transfer} for adapting new market knowledge and performing closed-set vendor verification on emerging (upcoming) vendors and markets. Finally, we evaluate the influence of knowledge transfer on our trained low-resource model against the zero-shot \citep{srivastava-etal-2018-zero} and transformers-based baselines. 

\section{Related Research}
\label{sec:related_research}
\paragraph{Vendor Verification - a supervised Authorship Attribution (AA) task:}
Researchers previously have utilized various NLP \citep{Ekambaranathan2018UsingST,10.1145/3292500.3330763, https://doi.org/10.48550/arxiv.2207.03477} and computer vision \citep{10.1145/3196494.3196529,DBLP:journals/corr/HeZRS15} techniques to identify and link vendors across Darknet markets. For example, in their research, \citet{10.1145/3308558.3313537} proposed uStyle-uID to leverage writing and photography styles to identify vendors in drug trafficking markets. Similarly, \citet{10.1145/3366423.3380263} proposed exploiting the multi-view learning paradigm and domain-specific knowledge to improve the cross-domain performance with both stylometric and location representation.

The Darknet ads consist of a product title and description, vendor name, price of the product, and occasionally some meta-data and images. While most of these details were enclosed in the ad's description, manual extraction of these features requires considerable labeling efforts. Therefore, we emphasize our research towards an end-to-end approach that only expects the advertisement's title and description to analyze the writing patterns for vendor verification and identification. Furthermore, since we perform multi-class classification over the text sequences of Darknet ads, we consider our approach similar to the AA task in NLP.

With the advances in NLP, there has been considerable research into the field of AA that has demonstrated the success of TFIDF-based clustering and classification techniques \citep{10.1145/3368567.3368572,Bozkurt2007AuthorshipA}, CNNs \citep{Rhodes2015AuthorAW,shrestha-etal-2017-convolutional}, RNNs \citep{8663814,jafariakinabad2019syntactic,10.1145/3325917.3325935}, and transformers architectures \citep{Fabien2020BertAAB,Ordoez2020WillLP,Uchendu2020AuthorshipAF}. Moreover, researchers have also observed a significant difference in language structure between Darknet and Surface net websites \citep{choshen-etal-2019-language, https://doi.org/10.48550/arxiv.2204.06885}. Therefore, exploring the application of authorship tasks on the Darknet language is crucial.

\paragraph{Vendor Identification; A Text Similarity task:} 
Text-similarity techniques are not new to the researchers in the field of AA \citep{10.1007/978-3-642-37256-8_38, castro-castro-etal-2015-authorship, 10.1007/s11192-018-2661-6, https://doi.org/10.48550/arxiv.1910.08144}. However, with the recent success of transformers \citep{reimers-gurevych-2019-sentence,Yang2019XLNetGA, jiang2022promptbert}, researchers are now investigating the application of semantically meaningful representations for paraphrasing detection \citep{9750249, 10.1007/978-3-030-78270-2_50, 9070636}, text summarization \citep{https://doi.org/10.48550/arxiv.1906.04165, CAI2022103999}, semantic parsing \citep{10.1007/978-3-030-32236-6_70, ferraro-suominen-2020-transformer}, question answering \citep{Yang_2019, 10.1145/3462757.3466102, https://doi.org/10.48550/arxiv.2108.11792}, and AA \citep{Fabien2020BertAAB, https://doi.org/10.48550/arxiv.2011.05864, CUSTODIO2021114866, uchendu-etal-2020-authorship}.

The recent developments in style representations \citep{hay-etal-2020-representation, zhu-jurgens-2021-idiosyncratic} have revealed a promising avenue to explore for the authorship verification task. In their research, \citet{wegmann-etal-2022-author} discovered that the success of these representations comes from their ability to represent style by latching on to spurious content correlations. Moreover, the authors suggest using content control in a contrastive setup to represent style better in a way that is more independent from content. In this research, we utilize a similar approach to extract the style representations from the advertisements of darknet vendors by passing it through a Transformer-based classifier pre-trained for a closed-set vendor verification task. Next, we use these representations to compute text similarity (cosine similarity) in the advertisements of different vendors. Despite our acknowledgment of the promises of using content control on style representations, this research focuses on establishing a baseline on Darknet markets. That being said, we intend to experiment with content control in our future experiments.  

\paragraph{Knowledge Adaption; A Transfer Learning task:} 
In their research, \citet{ruder2019neural} introduced transfer learning to extract knowledge from a source setting and transfer it to a target setting. Since then, many researchers have investigated the successful application of transfer learning on the cross-domain and topic AA task \citep{sapkota-etal-2014-cross, Barlas2021ATL}. Similar to the experiments in \citep{devlin2019bert, horne-etal-2020-grubert}, this work proposes utilizing knowledge transfer to adapt new market knowledge from the emerging Darknet vendors and markets. The transfer is applied using pre-trained style representations to train a computationally efficient BiGRU classifier for the closed-set vendor verification task. 

\section{Datasets}
\label{sec:datasets}
Many researchers have conducted similar experiments on scraped data from active Darknet markets. However, since law enforcement has seized and shut down these markets now, we could not reproduce the results nor get access to their data. Therefore, for reproducibility and future research purposes, we conduct our analyses on public datasets from Alphabay \citep{10.5555/3277203.3277279,inbookalphabay,alphabay}, Dreams, Traderoute, Valhalla, and Berlusconi \citep{Carr2019IntoTR,dreams}, Agora \citep{dnmArchives}, and Silk Road \citep{10.1145/2488388.2488408,silk-road1}  non-anonymous markets.\footnote{Hosted by \href{https://www.impactcybertrust.org/}{IMPACT cyber trust portal}}



\begin{table}[]
\centering
\resizebox{\columnwidth}{!}{%
\begin{tabular}{|l|l|l|l|}
\hline
\textbf{Task} &
  \textbf{Dataset} &
  \textbf{Ads.} &
  \textbf{Vendors} \\ \hline \hline
\multirow{4}{*}{\begin{tabular}[c]{@{}l@{}}\textbf{Baseline} / \\ \textbf{Supervised}\\ \textbf{Pre-Training}\end{tabular}} &
  Alphabay &
  100,429 &
  1,457 \\ \cline{2-4}
 & Dreams               & 93,586  & 1,422 \\ \cline{2-4}
 & Silk Road-1          & 78,681  & 1,392 \\ \cline{2-4}
 & \textbf{Alphabay-} & \textbf{272,696} & \textbf{4,271} \\
 & \textbf{Dreams-Silk} & \textbf{} & \textbf{} \\ \cline{2-4}  \hline \hline
\multirow{3}{*}{\begin{tabular}[c]{@{}l@{}}\textbf{Low-Resource}\\ \textbf{Supervised}\\ \textbf{Market Adaption} \end{tabular}} &
  Valhalla &
  2,175 &
  110 \\ \cline{2-4}
 & Berlusconi           & 1,437   & 84    \\ \cline{2-4}
 & \textbf{Valhalla-}  &\textbf{3,612}   & \textbf{194}   \\ 
  & \textbf{Berlusconi}  &   &   \\ \cline{2-4} \hline \hline
\multirow{3}{*}{\begin{tabular}[c]{@{}l@{}}\textbf{High-Resource}\\ \textbf{Supervised}\\ \textbf{Market Adaption} \end{tabular}} &
  Traderoute &
  19,952 &
  612 \\ \cline{2-4}
 & Agora                & 109,644 & 3,187 \\ \cline{2-4}
 &
  \begin{tabular}[c]{@{}l@{}}\textbf{Traderoute-}\\ \textbf{Agora}\end{tabular} &
  \textbf{129,586} &
  \textbf{3,799} \\ \cline{2-4} \hline
\end{tabular}%
}
\caption{Number of unique ads and vendor accounts per market.}
\label{tab:datasets}
\end{table}

\begin{minipage}[h]{\linewidth}
\includegraphics[ width=\linewidth]{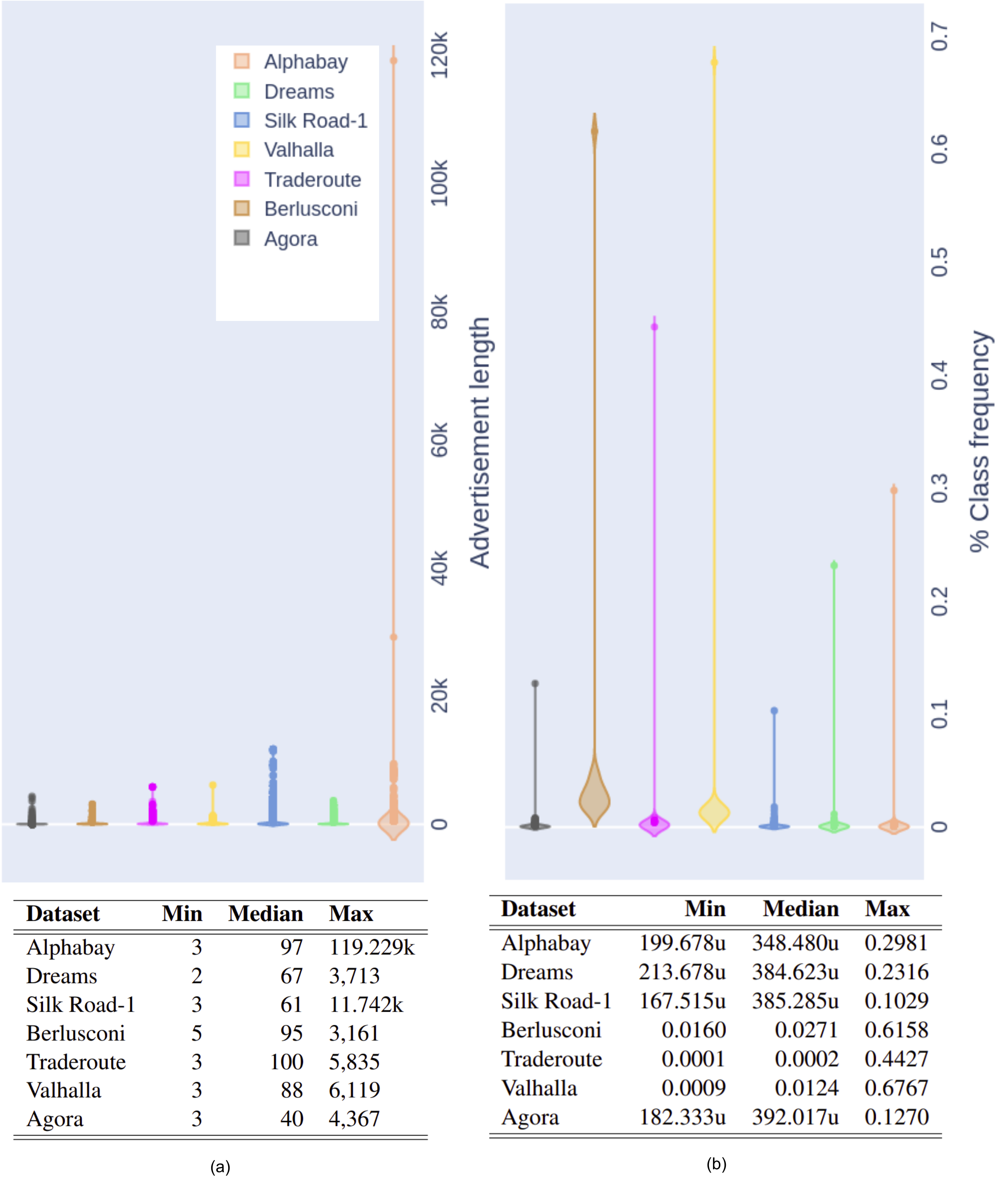}
\captionof{figure}{Distribution of (a) Token length per advertisement (b) Number of ads per vendor.\newline}
\label{fig:dataset_distribution}
\end{minipage}

\paragraph{Preprocessing:}Figure \ref{fig:dataset_distribution}(a) demonstrates the distribution of the number of tokens for all the input ads in our datasets. In a violin plot, the probability distribution is maximum around the median, and Table \ref{fig:dataset_distribution}(a) shows that the median for our chosen datasets is between 40 and 100. Therefore, to compare other baseline classifiers and transformers-based models fairly, we truncate our ads to the first 512 tokens. On the other hand, figure \ref{fig:dataset_distribution}(b) demonstrates a class imbalance in the number of ads per vendor account in our datasets. As can be seen, some markets are more imbalanced than others. Therefore, in contrast to earlier research emphasising the performance of the trained models on accuracy and micro-F1, we also evaluate our trained models on macro-F1, which weighs all classes equally. \newline

\noindent Table \ref{tab:datasets} illustrates the number of unique ads (input sequences) and vendor accounts per market.\footnote{In this research, market data refers to the ads and vendor accounts from a single Darknet market. On the other hand, a dataset refers to the combined data from two or more markets.} First, we merge the title and description of the ads using the "[SEP]" token to form the input sequences. Then, we drop all the duplicate ads for every vendor in our dataset. Most ads are in English, with a few exceptions where the vendors use multiple languages. We reason that the noise in the data roughly represents the unique writing style of individual vendors. For example, we found that the vendor "CaliforniaDreams420" refers to medicines as "medi...",  "SAPIOWAX" uses multiple "-" for newline, and "QualityKing" only uses uppercase letters in its ads. Therefore, any cleaning and processing will only be counter-productive. However, since we consider the vendor accounts as the gold labels for our classification task, we lower-cased all the vendor names to minimize the number of vendors in our datasets. In other words, we assume the vendors "agentq" and "AgentQ" to be the same entity. The table illustrates how we divide our datasets for supervised pre-training, Low-Resource, and High-Resource fine-tuning steps. Finally, we assign all the vendors with less than 20 ads to a new class label, "others," allowing our classifier to operate in a zero-shot setting. 

While we do not perform classification for vendors with less than 20 ads, we capture similarities in the ads for these vendors through our open-set vendor identification task. That said, the number 20 is not arbitrary and is established through experiments. We also experimented with the same setup by removing vendors with less than 5, 10, 20, 50, and 100 ads. The results demonstrate that our model requires at-least 20 ads from each vendor to perform the classification optimally. 

\section{Experiments}
Before running our experiments, we conduct a sanity check to evaluate the need for ML algorithms by examining the similarity in Darknet ads using \href{https://pypi.org/project/textdistance/}{textdistance}-based traditional stylometric approaches \cite{textdistance} (refer appendix \ref{app:traditional_stylometric}). Our analyses show that these traditional methods fail to identify vendors with dissimilar ads, indicating the need for sophisticated feature-extraction techniques. Furthermore, these approaches help us discard identical ads from further analysis.

\subsection{Closed-Set Vendor Verification Task}
\label{sec:baselines}
\paragraph{Architectural Baselines:} To verify the vendor migrants existing across multiple markets, we first train multiple classifiers to examine different writing styles in Darknet ads and establish a benchmark amongst various ML and neural network-based algorithms. Given the resources at our disposal, training models on the combined Alphabay, Dreams, and Silk Road datasets would be computationally expensive and time-consuming. Therefore, we first establish an architectural baseline by training (i) TFIDF-based statistical (Multinomial Naive Bayes, Logistic Regressor, Random Forest, SVMs, and MLP network), (ii) Bi-directional GRU with Fasttext embeddings \citep{10.1145/3325917.3325935}, CNNs over character n-grams \citep{shrestha-etal-2017-convolutional}, (iii) Pre-trained BERT-base-cased \citep{devlin2019bert}, RoBERTa-base \citep{liu2019roberta}, and a DistilBERT-base-cased \citep{sanh2019distilbert} sequence classifiers to identify 1,422 unique vendor accounts from 93,586 ads on the Dreams market.

\paragraph{Methodological Baselines:} We further establish a methodological baseline to investigate the influence of different training approaches on the combined Alphabay, Dreams, and Silk Road 1 dataset with 272,696 ads and 3,896 unique vendors. First, we train BERT-base-cased and uncased classifiers to investigate the influence of uppercase and lowercase patterns in ads on the model's performance. Second, we investigate if applying knowledge transfer from a BERT-cased model, trained on the Darknet ads for the language task, improves the classification performance. In this research, we refer to the trained language model as \textit{DarkBERT-LM} and the classifier as \textit{DarkBERT-classifier}. In another study, \citet{houlsby2019parameterefficient} suggests that rather than updating the weights of the pre-trained model, it is much more efficient to stitch adapter layers and update them while keeping the pre-trained model frozen. Therefore, we finally train a BERT-cased classifier with adapter layers (aka \textit{Adapter-BERT}) and compute its performance.

\subsection{Open-Set Vendor Identification Task}
\label{exp:representational_similarity}
In their research, \citep{kornblith2019similarity, phang2021fine} proposed \textit{Centered Kernel Alignment (\href{https://pypi.org/project/torch-cka/}{CKA})} as a similarity metric to reliably compute correspondences between representations in networks trained from different initializations. In this research, we compute CKA similarity between the representational layers of our trained classifier and an available pre-trained checkpoint (not trained on Darknet data). Finally, we examine the least similar layers, i.e., the layers that changed most during training and have a low CKA similarity, to extract semantically-meaningful style representations from the ads of Darknet markets. \footnote{Algorithm-\ref{algo:cka} in Appendix \ref{app:psuedo_code} demonstrates the pseudo-code for computing CKA similarity across layers of our trained classifier and an available pre-trained checkpoint.}

Similar to \citet{reimers2019sentencebert}, we compute the similarity between two vendors by computing cosine-similarity between the extracted style representations in their ads. Then, assigning one of the vendors as the parent vendor, we repeat the process for all the other vendors in our dataset. However, cosine distance represents a linear space with all dimensions weighted equally. Therefore, \citet{xiao2018bertservice} suggests that the emphasis be on the rank and not the absolute value representing the similarity between the two vendors. Besides, vendors on Darknet advertise their products across various categories. For two vendors, A, and B, selling their products under multiple categories, the cosine similarity between their ads would be low by default. Therefore, instead of comparing ads across similar trade categories (which requires labeling efforts and is counterproductive to our research), we propose normalized similarity ($sim_{norm}$) as a measure of cosine similarity ($sim$) in ads between two vendors, w.r.t. to the self-similarity ($sim_{self}$) in their ads through the equation below:

\[sim_{norm}=2*\frac{sim(A, B)}{sim_{self}(A,A)+sim_{self}(B, B)}\]

\subsection{Low-Resource Market Adaption Task}
\label{exp:knowledge-transfer}
To verify the vendor migrants from emerging markets, we conduct experiments on an LR dataset, i.e., Valhalla-Berlusconi, with 3,612 ads and 194 vendors. First, we extract the style representations from the "[CLS]" token of the pre-trained classifier (Section \ref{sec:baselines}) for all the ads in our LR dataset. Then, following \citep{devlin2019bert}, we apply knowledge transfer from the pre-trained classifier to a two-layer bidirectional GRU classifier by initializing it with the extracted style representations. The Bi-GRU classifier is then fine-tuned to adapt new market knowledge and verify the migrants across the LR dataset. Our research refers to this as the \textit{transfer-BiGRU} model. Performing knowledge transfer helps our existing classifier to evolve with emerging vendor and Darknet market data. During the evaluation, we compare the performance of our transfer-BiGRU against BERT-base-cased and two-layer BiGRU (with fasttext embeddings) classifiers (aka end-to-end baselines) when trained from scratch on the LR dataset. Finally, we also evaluate the zero-shot performance of our architectural and methodological classifiers (aka zero-shot baselines) against the transfer-BiGRU for the closed-set vendor verification task.

\section{Results}
\subsection{Open-Set Vendor Verification Task}
\label{result:establishing-baseline}

\paragraph{Architectural Baselines:}Table \ref{tab:architectural_baselines} presents the performance of our architectural baselines evaluated on the Dreams market. Amongst all the statistical models, we found a Multilayer Perceptron (MLP) with bigram TF-IDF features to perform the best. While conventional neural networks such as character-based CNN and Bidirectional GRU with fasttext embeddings performed better than the statistical models, we noted a considerable increase in performance with the transformers-based architecture on our datasets. To our surprise, the RoBERTa-base model underperformed compared to the BERT-base-cased architecture. Although we propose to leverage writing styles to identify various vendors, the Darknet markets are intentionally designed with random noise to foil any automated system. Furthermore, since RoBERTa-tokenizer works on "byte-level BPE," we believe the trained model did not have enough data to learn these features. Consequently, we establish the trained BERT-cased classifier on the Dreams market as the benchmark classifier of our architectural baselines.

\begin{table}[htp]
\centering
\resizebox{\columnwidth}{!}{%
\begin{tabular}{|c|llll|}
\hline
\multicolumn{1}{|l|}{\textbf{Data}} & \multicolumn{1}{l|}{\textbf{Models}} & \multicolumn{1}{l|}{\textbf{Accuracy}} & \multicolumn{1}{l|}{\textbf{Micro-F1}} & \textbf{Macro-F1} \\ \hline
\multirow{13}{*}{\begin{tabular}[c]{@{}c@{}}Dreams\\ \\ Market\end{tabular}} & \multicolumn{4}{c|}{Statistical Models} \\ \cline{2-5} 
 & \multicolumn{1}{l|}{\begin{tabular}[c]{@{}l@{}}Multinomial \\ Naive Bayes\end{tabular}} & \multicolumn{1}{l|}{0.0183} & \multicolumn{1}{l|}{0.0144} & 0.0059 \\ \cline{2-5} 
 & \multicolumn{1}{l|}{Random Forest} & \multicolumn{1}{l|}{0.0102} & \multicolumn{1}{l|}{0.1093} & 0.0449 \\ \cline{2-5} 
 & \multicolumn{1}{l|}{\begin{tabular}[c]{@{}l@{}}Logistic \\ Regression\end{tabular}} & \multicolumn{1}{l|}{0.0045} & \multicolumn{1}{l|}{0.0090} & 0.0037 \\ \cline{2-5} 
 & \multicolumn{1}{l|}{SVM} & \multicolumn{1}{l|}{0.2480} & \multicolumn{1}{l|}{0.3974} & 0.3703 \\ \cline{2-5} 
 & \multicolumn{4}{c|}{Neural Networks} \\ \cline{2-5} 
 & \multicolumn{1}{l|}{MLP} & \multicolumn{1}{l|}{0.6614} & \multicolumn{1}{l|}{0.6603} & 0.6594 \\ \cline{2-5} 
 & \multicolumn{1}{l|}{Character-CNN} & \multicolumn{1}{l|}{0.7266} & \multicolumn{1}{l|}{0.7256} & 0.7248 \\ \cline{2-5} 
 & \multicolumn{1}{l|}{BiGRU-Fasttext} & \multicolumn{1}{l|}{0.7374} & \multicolumn{1}{l|}{0.7415} & 0.7360 \\ \cline{2-5} 
 & \multicolumn{4}{c|}{Transformers Networks} \\ \cline{2-5} 
 & \multicolumn{1}{l|}{\textbf{BERT-cased}} & \multicolumn{1}{l|}{\textbf{0.8978}} & \multicolumn{1}{l|}{\textbf{0.8978}} & \textbf{0.9002} \\ \cline{2-5} 
 & \multicolumn{1}{l|}{DistilBERT-cased} & \multicolumn{1}{l|}{0.8886} & \multicolumn{1}{l|}{0.8885} & 0.8889 \\ \cline{2-5} 
 & \multicolumn{1}{l|}{RoBERTa-base} & \multicolumn{1}{l|}{0.8776} & \multicolumn{1}{l|}{0.8797} & 0.8736 \\ \hline
\end{tabular}%
}
\caption{Performance of architectural baselines on the Dreams market.}
\label{tab:architectural_baselines}
\bigskip

\resizebox{\columnwidth}{!}{%
\begin{tabular}{|l|l|l|l|l|}
\hline
\textbf{Data} & \textbf{Models} & \textbf{Accuracy} & \textbf{Micro-F1} & \textbf{Macro-F1} \\ \hline
\multirow{4}{*}{\begin{tabular}[c]{@{}l@{}}Alphabay-\\ Dreams-\\ Silk\\ Dataset\end{tabular}} & BERT-uncased & 0.8947 & 0.8939 & 0.8768 \\ \cline{2-5} 
 & \textbf{BERT-cased} & \textbf{0.9046} & \textbf{0.9066} & \textbf{0.9013} \\ \cline{2-5} 
 & \begin{tabular}[c]{@{}l@{}}DarkBERT-\\ Classifier\end{tabular} & 0.9000 & 0.9090 & 0.9073 \\ \cline{2-5} 
 & Adapter BERT & 0.8398 & 0.8330 & 0.8188 \\ \hline
\end{tabular}%
}
\caption{Performance of methodological baselines on the combined Alphabay-Dreams-Silk dataset.}
\label{tab:methodological_baselines}
\end{table}

\paragraph{Methodological Baselines:} Table \ref{tab:methodological_baselines} illustrates the performance of our methodological baselines evaluated on the combined Alphabay-Dreams-Silk Road-1 test dataset. Our first experiment investigates the influence of writing style, i.e., lowercase and uppercase patterns, on the classification task. As can be seen, the BERT-cased classifier outperforms the uncased classifier by a reasonable margin (Approx. 3\% on 3,896 class labels). We believe that the increment in performance comes from adding upper and lowercase patterns during training. Next, we experiment with continued pre-training of the DarkBERT-LM on the ads for the language task \footnote{Pre-training BERT for a language task is highly resource-intensive. Unfortunately, we did not have the resources to continue the pre-training until the convergence and only trained our model for 20 epochs.} to achieve a test perplexity of 2.07. In comparison to the BERT-cased classifier, we observe a minor increase in the performance of the finetuned DarkBERT-Classifier. However, we reason that such a minor increase is not worth all the training. Furthermore, the low performance of the DarkBERT-LM depicts the unpredictable and noisy lingo used by Darknet vendors in their ads. We also suspect that further pre-training our models on an extensive dataset can help the baseline improve its performance. Finally, the Adapter BERT also underperforms compared to the vanilla BERT-cased classifier. Consequently, we establish the BERT-cased architecture trained on the closed-set vendor verification task as the benchmark classifier for the Alphabay-Dreams-Silk Road Darknet dataset.

\subsection{Open-Set Vendor Identification Task}
\definecolor{light_green}{rgb}{0.75, 0.95, 0.77}
\definecolor{light_red}{rgb}{0.96, 0.71, 0.69}
\definecolor{light_blue}{rgb}{0.82, 0.70, 0.87}

\begin{table}[h]
\centering
\begin{minipage}{0.49\textwidth}
	\centering
	\includegraphics[width=\linewidth]{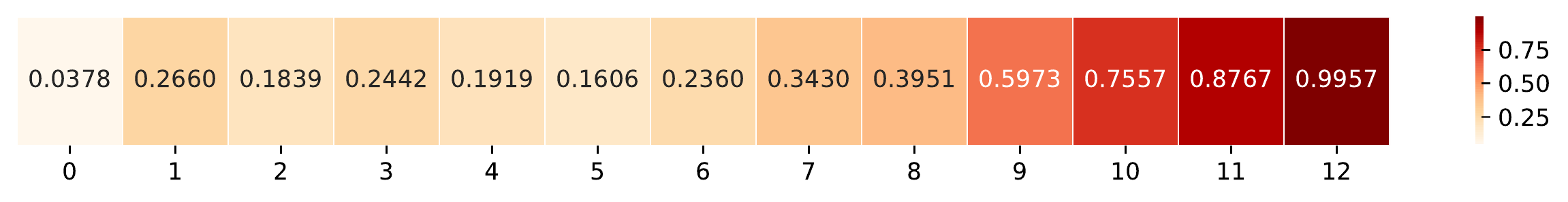}
	\captionof{figure}{\label{fig:cka_similarity}CKA distance between layers of the BERT-cased methodological classifier, compared before and after being trained on the Alphabay-Dreams-Silk dataset.}
	\bigskip
	\includegraphics[width=\linewidth]{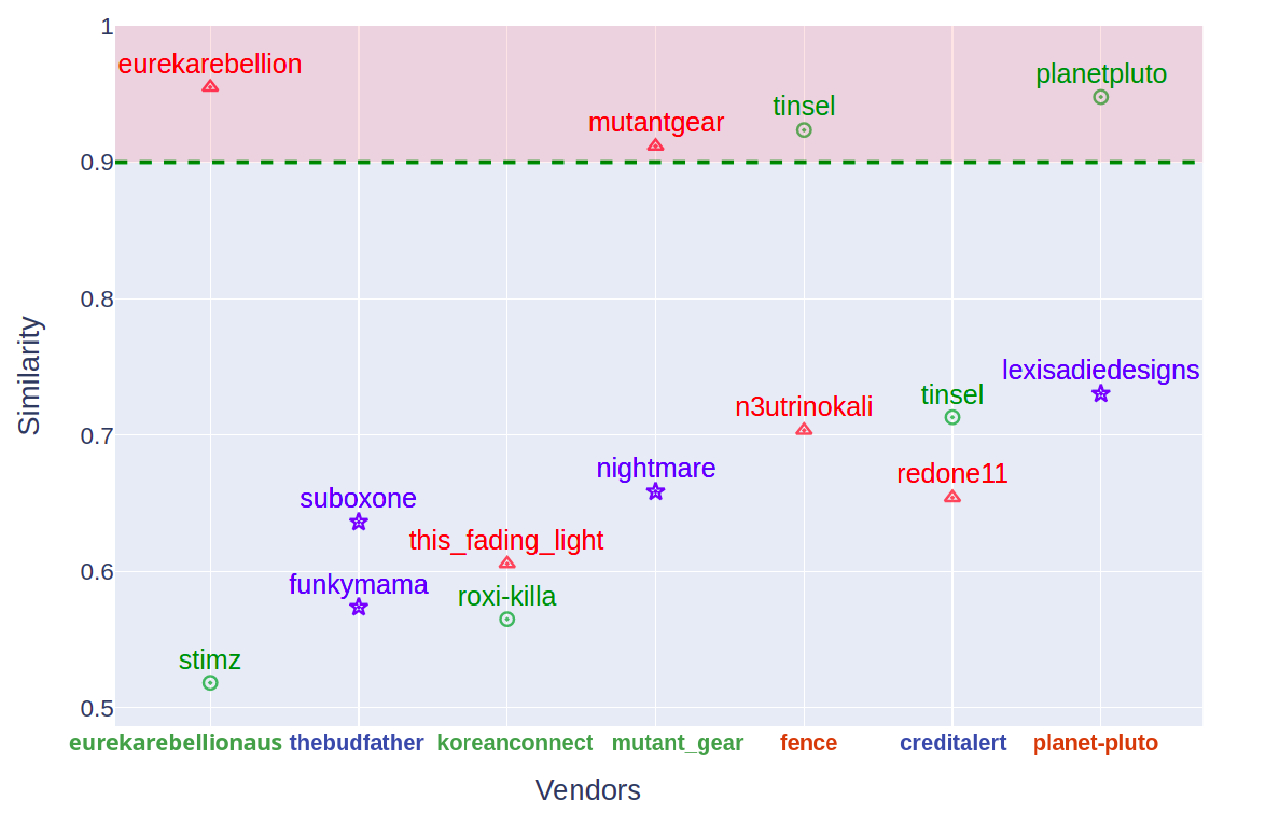}
	\captionof{figure}{\label{fig:cosine_similarity} Scatter plot between parent-vendors (on the x-axis) and their potential aliases (scatter points on y-axis) from \colorbox{light_red}{Alphabay}, \colorbox{light_green}{Dreams}, and \colorbox{light_blue}{Silk Road-1} markets.}
\end{minipage}
\end{table}

\noindent Figure \ref{fig:cka_similarity} reveals a high CKA distance, i.e., low CKA similarity, between the representations for the last four layers of our BERT-cased classifier. Therefore, extracting information from the weighted sum of the final four layers provides the most meaningful style representations for our ads in the Alphabay-Dreams-Silk dataset. As described in section \ref{exp:representational_similarity}, we use these style representations to compute the cosine similarity between vendor ads. Figure \ref{fig:cosine_similarity} displays some randomly selected parent vendors on the x-axis and their two potential aliases (scatter points) with a similarity score in their ads on the y-axis. \footnote{We generate the scatter plot using \href{https://plotly.com/python/}{Plotly}, which allows us to zoom infinitely for any vendor. However, we only show the chosen vendors with two potential aliases for better clarity and visibility.} Our analysis indicates "eurekarebellionaus" and "eurekarebellion," "mutant\textunderscore gear" and "mutantgear", "fence" and "tinsel," and "planet-pluto" and "planetpluto" have very high similarity in their ads and can be from the same vendor. The higher the similarity, the more likely it is for two vendors to be the same entity. \footnote{As mentioned, Darknet vendors often create aliases to hide from Law Enforcement Agencies (LEA). However, since numerous vendors appear on Darknet markets yearly, it becomes difficult for law enforcement to manually link these aliases to a parent vendor. The unavailability of
ground truth poses a challenge in evaluating the existence of these aliases in our datasets. Therefore, we cannot confidently comment upon the accuracy of our similarity-based analyses without the qualitative case study. We encourage LEA not solely to rely on these similarities but use them as a starting point for their manual investigations. Furthermore, we strongly discourage LEA from abiding by these analyses as evidence for investigation or prosecution. The sole purpose of this research is to help LEA bring meaning to the online Darknet market data.} For a better visibility, these vendors are highlighted inside the red box of our scatter plot. 

\begin{table}[]
\resizebox{\columnwidth}{!}{%
\begin{tabular}{|l|l|l|l|}
\hline
\textbf{} & \textbf{Parent Vendor} & \textbf{Alias / Copycat} & \textbf{Similarity} \\ \hline
\multirow{5}{*}{\begin{tabular}[c]{@{}l@{}}High\\ (potential\\ aliases)\end{tabular}} & houseofdank & houseofdank2.0 & 0.9844 \\ \cline{2-4} 
 & incorporated & incorporatedv2 & 0.9769 \\ \cline{2-4} 
 & castro6969 & castro69696 & 0.9541 \\ \cline{2-4} 
 & thewizard & thewizzardnl & 0.9480 \\ \cline{2-4} 
 & europills & europills2 & 0.9467 \\ \hline \hline
\multirow{5}{*}{\begin{tabular}[c]{@{}l@{}}Low\\ (potential\\ Copycats)\end{tabular}} & topgear & topgear69 & 0.0367 \\ \cline{2-4} 
 & dutchpirates & dutchpiratesshop & -0.1015 \\ \cline{2-4} 
 & whitey & whiteyford & -0.1410 \\ \cline{2-4} 
 & g3cko & gecko & -0.2292 \\ \cline{2-4} 
 & aussieimportpills & aussieimportpillsv2 & -0.2560 \\ \hline
\end{tabular}%
}
\caption{Normalized similarity between parent vendors and their potential aliases/copycats aligned in decreasing order.}
\label{tab:rep_similarity}
\end{table}

\noindent Often, vendor aliases have similar-looking vendor handles to have recognition and a monopoly over their business. While most similar-looking accounts can be detected using string-based matching techniques like \href{https://pypi.org/project/string-grouper/}{string$\_$grouper} \cite{stringgrouper}, our experiments reveal the existence of copycats with very different writing styles and low similarity in their ads. For example, our experiments uncovered that only about 24\% of similar-looking vendor-alias pairs in the Alphabay-Dreams-Silk dataset have a similarity score of 0.7 or above in their ads. Table \ref{tab:rep_similarity} illustrates the similarity in ads between 10 such parent vendors and their likely aliases or copycats. Finally, we believe our experiments can also help law enforcement uncover potential vendor-alias pairs with completely unrelated vendor names, ex: "fence" and "tinsel" (see figure \ref{fig:cosine_similarity}), but a high similarity between their ads.

\subsection{Low Resource Market Adaption Task}
\label{result:knowledge-transfer}
To set the Zero-Shot baselines, we first use the established BERT-cased architectural and methodological classifiers to perform zero-shot vendor verification on the LR dataset, Valhalla-Berlusconi. Since the emerging LR dataset has new vendors, we assign all these new vendor accounts to the class label "others." However, since the macro-F1 score is computed for the unweighted arithmetic mean of F1 for all class labels, the absence of previously existing vendors in the LR emerging market leads us to unreliable macro-F1 results. Consequently, we emphasize the performance of our Zero-Shot baselines on the micro-F1 score. The baselines exhibit promising performance with a micro-F1 of 0.7702 and 0.7388 despite not being trained on LR data. The decrease in macro-F1 performance from architectural to methodological baseline is due to an increase in vendor accounts from 1,442 in the Dreams market to 3,896 in the Alphabay-Dreams-Silk Road dataset.

\begin{table}[htp]
\centering
\resizebox{\columnwidth}{!}{%
\begin{tabular}{|llll|}
\hline
\multicolumn{1}{|l|}{\textbf{Models}} & \multicolumn{1}{l|}{\textbf{Layer}} & \multicolumn{1}{l|}{\textbf{Micro-F1}} & \textbf{Macro-F1} \\ \hline
\multicolumn{4}{|c|}{\textit{Zero-Shot Baselines}} \\ \hline
\multicolumn{1}{|l|}{Architectural} & \multicolumn{1}{c|}{-} & \multicolumn{1}{l|}{0.7702} & 0.2927 \\ \hline
\multicolumn{1}{|l|}{Methodological} & \multicolumn{1}{c|}{-} & \multicolumn{1}{l|}{0.7388} & 0.2401 \\ \hline
\multicolumn{4}{|c|}{End-to-End Baselines} \\ \hline
\multicolumn{1}{|l|}{\textbf{BERT-cased}} & \multicolumn{1}{c|}{-} & \multicolumn{1}{l|}{\textbf{0.8987}} & \textbf{0.8148} \\ \hline
\multicolumn{1}{|l|}{BiGRU-Fasttext} & \multicolumn{1}{c|}{-} & \multicolumn{1}{l|}{0.7797} & 0.6957 \\ \hline
\multicolumn{4}{|c|}{\textit{Transfer Baselines}} \\ \hline
\multicolumn{1}{|l|}{\multirow{5}{*}{\begin{tabular}[c]{@{}l@{}}\textbf{Transfer-}\\ \textbf{BiGRU}\end{tabular}}} & \multicolumn{1}{l|}{Embedding} & \multicolumn{1}{l|}{0.7653} & 0.6408 \\ \cline{2-4} 
\multicolumn{1}{|l|}{} & \multicolumn{1}{l|}{Last} & \multicolumn{1}{l|}{0.8590} & 0.7809 \\ \cline{2-4} 
\multicolumn{1}{|l|}{} & \multicolumn{1}{l|}{Second-to-Last} & \multicolumn{1}{l|}{0.8951} & 0.7884 \\ \cline{2-4} 
\multicolumn{1}{|l|}{} & \multicolumn{1}{l|}{Weighted Sum All 12} & \multicolumn{1}{l|}{0.8928} & 0.7837 \\ \cline{2-4} 
\multicolumn{1}{|l|}{} & \multicolumn{1}{l|}{\textbf{Weighted Sum Last 4}} & \multicolumn{1}{l|}{\textbf{\textbf{0.8946}}} & \textbf{0.8132} \\ \hline
\end{tabular}%
}
\caption{Performance of Zero-Shot, End-to-End, and Transfer baselines on the Valhalla-Berlusconi dataset.}
\label{tab:lr_baseline}
\resizebox{\columnwidth}{!}{%
\begin{tabular}{|c|c|c|c|}
\hline
\textbf{GPU} & \textbf{Models} & \multicolumn{1}{l|}{\textbf{\begin{tabular}[c]{@{}l@{}}Trainable\\ parameters\end{tabular}}} & \multicolumn{1}{l|}{\textbf{\begin{tabular}[c]{@{}l@{}}Training Time\\   (Hrs:Mins)\end{tabular}}} \\ \hline
\multirow{3}{*}{\begin{tabular}[c]{@{}c@{}}Tesla-\\ V100\\ (32 GB)\end{tabular}} & BERT-cased & 110M & 0:54 \\ \cline{2-4} 
 & BiGRU-Fasttext & 13M & 0:12 \\ \cline{2-4} 
 & Transfer-BiGRU & 24M & 0:32 \\ \hline
\begin{tabular}[c]{@{}c@{}}GE-MX110\\ (2 GB)\end{tabular} & Transfer-BiGRU & 24M & 2:40 \\ \hline
\end{tabular}%
}
\caption{Computational details of trained classifiers on the LR, Valhalla-Berlusconi, dataset.}
\label{tab:lr_compuations}
\end{table}

Furthermore, we also train a BERT-cased and a BiGRU classifier with fasttext embeddings from scratch to adapt new market knowledge and vendors from the emerging LR dataset. As illustrated in table \ref{tab:lr_baseline}, compared to the Zero-Shot baselines, the End-to-End baselines show a significant increase in performance in both micro-F1 and macro-F1 scores. Finally, following \citep{devlin2019bert}, we perform knowledge transfer by extracting the style representations from multiple layers of the BERT-cased methodological classifier and using them to initialize the BiGRU before the classification layer. Table \ref{tab:lr_baseline} demonstrates that when initialized with the weighted sum of the last four layers, the transfer-BiGRU classifier benefits most from the knowledge transfer and performs comparably to the End-to-End BERT-cased classifier on the emerging LR dataset. Consequently, we establish the transfer-BiGRU architecture trained on the closed-set vendor verification task as the benchmark classifier for the LR, Valhalla-Berlusconi dataset. \footnote{We also test the performance of our baselines on an emerging High-Resource (HR) dataset, Traderoute-Agora. Results in the appendix table \ref{tab:hr_baseline} show that the transfer-BiGRU model underperforms compared to the End-to-End BERT-cased classifier. In other words, applying knowledge transfer on emerging HR markets does not yield the best performance. Please refer to section \ref{app:hr_baseline} in the appendix for more details.}

Finally, Table \ref{tab:lr_compuations} reflects upon the computational aspects of the trained models by comparing the number of trainable parameters and training time for classifiers trained on the LR dataset. As can be seen, compared to the BERT-cased, our transfer-BiGRU classifier is carbon-efficient (refer to appendix \ref{app:broader_impact}), has 78\% less trainable parameters, and takes approximately half the training time. Furthermore, we also show the training feasibility of our transfer-BiGRU on a low-end graphic card, GeForce-MX110, with 2 GB of GPU memory. Thus, our low-compute transfer-BiGRU classifier can significantly help law enforcement scale our approach to emerging markets without significant performance loss.

\section{Error Analysis}
\begin{table}[h]
\centering
\resizebox{\columnwidth}{!}{%
\begin{tabular}{|l|l|l|l|}
\hline \textbf{Vendor} & \textbf{Pred} & \textbf{Text A} & \textbf{Text B} \\ \hline
house & TP & ** 1 Lb of & ** 1 oz of \\
ofdank & & Sour [DRUG1] & Greenhouse [drug1] \\
& & (Greenhouse) **& greenhouse grown ** \\ \hline
house & TP & ** 1 OZ of & ** 1 Lb of \\
ofdank2.0 & & [drug2] Greehouse & [drug2] Greehouse \\
& & grown ** & grown ** \\ \hline
appleinc & FP & 10 x €50 euro & 5 x \$100 DOLLAR \\
 & & COUNTERFEIT & COUNTERFEIT \\
 & & notes (Very Good & STRIP high quality \\
 & & Quality) & bills \\ \hline 
canadian & FP & [usa to usa] [drug3] & [usa to usa] 30 pills \\
pharmacy & & 80mg just 19.99 & [drug3] 100mg 19.99 \\ 
 & & bucks per pill only & usd ultram \\ \hline
\end{tabular}
}
\caption{\label{tab:error_analysis1} Qualitative analysis of BERT-cased classifier (trained on Alphabay-Dreams-Silk Road Dataset) for True Positives (TP) and False Positives (FP) predictions.}
\end{table}

\noindent To better understand the strengths and weaknesses of our trained models, we perform qualitative analysis on the predictions of the BERT-cased classifier (trained on the Alphabay-Dreams-Silk Road Dataset) in Table \ref{tab:error_analysis1}. Note that we only display the title of these advertisements due to space constraints and visibility reasons. As can be seen in the first two examples, our trained classifier can recognize many patterns in the ads, such as "**," "[DRUG1]", "[drug1]", and "greenhouse gown,." The first two examples also show how similar the advertisements are between the vendors "houseofdank" and "houseofdank2.0". This is also indicated by the high similarity in the advertisements of the two vendors (refer \ref{tab:rep_similarity}). Finally, the next two examples in the Table below indicate the cases of false positives. As can be seen, here, the network is confusing between vocabulary such as "COUNTERFEIT," "quality," "supernotes," source and destination locations, [drug3], and the price of the product.

Furthermore, we inspect cases where our trained BERT-cased classifier fails, but the transfer-GRU classifier succeeds after knowledge transfer. Table \ref{tab:error_analysis2} demonstrates vendor advertisements where the writing style between advertisements changed drastically between the Alphabay-Dreams-Silk Road and Valhalla-Berlusconi datasets. Consequently, our BERT-cased classifier fails to verify vendors from the Valhalla-Berlusconi dataset in the zero-shot setting. Finally, after applying knowledge transfer and fine-tuning our transfer-BiGRU model, the model quickly adapts to the new writing styles from these vendor advertisements.

\begin{table}[h]
\centering
\resizebox{\columnwidth}{!}{%
\begin{tabular}{|l|l|l|}
\hline \textbf{Vendor} & \textbf{Alphabay-Dreams-} & \textbf{Valhalla-} \\ 
 & \textbf{Silk Road} & \textbf{Berlusconi} \\ \hline

cannacorner & [drug1] 3.5g & 7g [drug1] \\
 & ---|MERCEDES| & ---|lambo| \\ \hline
 
medicalznl & 5 GRAMS & 2.5 grams - \\
 & COLOMBIAN [DRUGX] & colombian [drugx] \\ 
  & 93\% + FREE SHIPPING & 90+\% pure uncut \\ \hline
  
color & Credit Cards Can Be & lasted update credit cards \\ 
 & Without Security Code & in this file. \\ \hline
\end{tabular}
}
\caption{\label{tab:error_analysis2} Qualitative analysis of transfer-BiGRU classifier (trained on Valhalla-Berlusconi Dataset) for True Positives (TP) and False Positives (FP) predictions.}
\end{table}

\section{Discussion and Future Work}
We discuss details about additional experiments and the training setup in appendix sections \ref{app:additional_exp} and \ref{appendix:infrastructure}, respectively. 

In the future, we plan to work on the assumptions and limitations indicated in appendix sections \ref{app:assumptions} and \ref{app:limitations} by investigating content-control contrastive learning approaches \citep{wegmann-etal-2022-author} to perform vendor verification and identification on existing and emerging Darknet datasets. 

\section{Conclusion}
This research presents an NLP-based vendor verification and identification approach, VendorLink, for law enforcement to verify, identify, and link vendor migrants and potential aliases on the existing and emerging Darknet markets. In this work, we first perform supervised pre-training to adapt Darknet market knowledge and establish a BERT-cased classifier to verify existing vendor migrants between markets in a closed-set environment. Then, we extract the style representations from the trained BERT-cased classifier to compute the text similarity in vendor ads in an open-set environment and link vendors to their potential aliases. Finally, we adapt new market knowledge by employing knowledge transfer from the trained BERT-cased classifier to a low-compute-resource BiGRU classifier and perform closed-set vendor verification on the emerging LR markets. Through our experiments, we uncover (i) 15 migrants and 71 potential aliases in the Alphabay-Dreams-Silk dataset, (ii) 17 migrants and 3 potential aliases in the Valhalla-Berlusconi dataset, and (iii) 75 migrants and 10 potential aliases in the Traderoute-Agora dataset with a cosine similarity of 0.8 and above, between the ads of vendors and their potential aliases. 

\section{Broader Impact}
\label{app:broader_impact}
This section discusses mandatory data collection protocols, ethical considerations, potential risks, and legal, societal, and environmental impacts.

\paragraph{Data Collection Protocol:} Ethical concerns associated with web scraping do not apply to our research as the online darknet data used is requested through a signed Memorandum of Agreement (MoA) with \href{https://www.impactcybertrust.org/}{IMPACT Cyber Trust portal} (ICC). As a result, the data is freely available, legally collected, and distributed for large-scale cybersecurity analytics, allowing researchers to advance the state-of-the-art cyber-risk R\&D and decision support.

\paragraph{Legal Impact:} This research emphasizes bringing structure and meaning to the massively available online data on Darknet markets for LEA. While we can not predict whether our research will impact the LEA process, the intent is to identify potential connections between vendors of illegal goods and present LEA with a broader information base for their internal processes. Please note that at no point do we claim to provide pieces of evidence necessary for prosecuting any criminal.

\paragraph{Ethical and Privacy Considerations:} We acknowledge that using vendor names in our study could potentially be exploited and identified as a privacy concern. However, after going through a Data Privacy Impact Assessment (DPIA) at our institution, the committee concluded that the vendor names used in this study are pseudonyms and do not reflect any individual's identity. Furthermore, research suggests that the lifespan of Darknet vendors and marketplaces is between a few months and a couple of years. \citep{9583681, 6ab479ad433946619944578604002da6, world_drug_report_2020}. Since the market ads in our datasets span between 2011-2018, the likelihood of any vendor's existence with the same user name is very low. Finally, under article 6, \href{https://gdpr-info.eu/art-6-gdpr/}{Lawfulness of processing}, the GDPR clause suggests that the processing of personal data is lawful as long as the task is carried out in the public interest. Given the nature of illegal activities on the Darknet and despite all its potential risks, we believe that our research can potentially benefit LEA and save human lives. That said, while using vendor names in our analyses promotes transparency and reproducibility amongst the readers, we encourage these vendors to reach out to us in case of any concerns. In such circumstances, we take complete responsibility for taking immediate action and removing their information from our research.

\paragraph{Societal Impact and Potential Risk:}
In their research, \citet{dark-side-social-analytics} described the dark side of authorship studies and social media analytics for target-based recommendation systems and employee, political, medical, gender, demographic, and racial profiling. While our approach can lend itself to abuses, we find it unlikely for anyone to exploit our research as it is given the extreme difference in the language between the Darknet and surface web websites \citep{choshen-etal-2019-language}. That said, we acknowledge the possibility of privacy infringement outside criminal markets to match user activity across public platforms. For instance, ill-intentioned third parties and organizations could use our research to circumvent an individual's identity on public social media platforms. Therefore, we encourage our readers to be aware of the ethical duality while using our research to develop authorship technologies inside and outside cybersecurity scenarios.

\paragraph{Environmental Impact:} Keeping in mind that not all LEA have the resources to train computationally expensive architectures, we investigate utilizing knowledge transfer to train low-compute-resource models in this research. As a result, our transfer-BiGRU classifier has a carbon efficiency of 0.07 kgCO$_2$eq/kWh and 2.25 kgCO$_2$eq/kWh as opposed to the BERT-cased classifier with a carbon efficiency of 0.12 kgCO$_2$eq/kWh and 4.21 kgCO$_2$eq/kWh on the Vallhalla-Berlusconi and Traderoute-Agora datasets, respectively. These estimations were conducted on Tesla V100-SXM2-32GB (TDP of 300W) using the \href{https://mlco2.github.io/impact#compute}{MachineLearning Impact calculator} presented in \citep{lacoste2019quantifying}. In other words, this research demonstrates that applying knowledge transfer from existing to emerging markets can help law enforcement train low-compute-resource models with high performance, faster training time, and lesser carbon footprint. 

\section{Acknowledgement}
This research is supported by the Sector Plan Digital Legal Studies of the Dutch Ministry of Education, Culture, and Science and Cora4NLP project, funded by the German Federal Ministry of Education and Research (BMBF) under funding code 01IW20010. Finally, the experiments were made possible using the Data Science Research Infrastructure (DSRI) hosted at Maastricht University. 

\bibliography{anthology,custom}
\bibliographystyle{acl_natbib}

\appendix
\section{Appendix}
\label{sec:appendix}
\subsection{Additional Experiments}
\label{app:additional_exp}

\subsubsection{Sanity Check: stylometric approaches}
\label{app:traditional_stylometric}

\noindent As a sanity check, we investigate the need for ML algorithms by examining if traditional stylometric approaches can identify writing patterns in Darknet ads. Since languages are represented by characters, tokens, and sentence-level elements, we compute string, token, and sequence-based similarities between ads using the Damerau-Levenshtein distance, Jaccard Index, and Ratcliff-Obershelp pattern recognition technique from \href{https://pypi.org/project/textdistance/}{textdistance}. We define the similarity between two vendor ads as the average of the above three metrics. For a vendor with multiple ads, say vendor A, we compute average similarity as the mean of similarities between all their ads. Similarly, for vendor B, existing across multiple markets, we take all the ads from market X and compute their similarity with ads of market Y (one at a time). Finally, we compute the average similarity as the mean of similarities between the ads for vendor B across all markets. 

Figure \ref{fig:style_similarity} demonstrates the performance of traditional stylometric approaches on a box plot. The plot represents the average similarity distribution and its skewness within the ads of Alphabay-Alphabay, Dreams-Dreams, Silk Road-Silk Road and across Alphabay-Dreams, Dreams-Silk Road, and Alphabay-Silk Road markets. As can be seen, most ads have an average similarity below 0.20. While there are outliers with higher similarities, only one vendor, "cyanspore", has a similarity score of 1.0 for the Alphabay-Dreams and Dreams-Silk datasets. Since the ads from this vendor are exactly similar, we remove them from all our further analyses.

\resizebox{\columnwidth}{!}{%
\begin{minipage}{\columnwidth}
\includegraphics[width=\columnwidth]{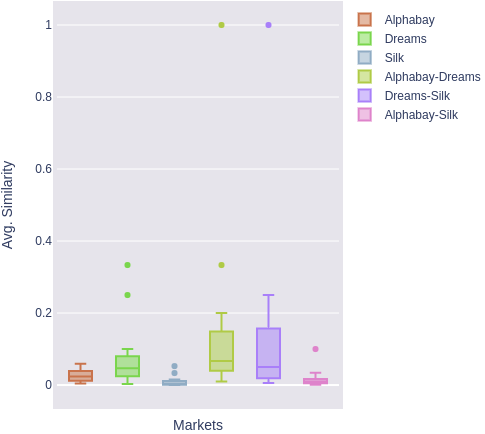}
\captionof{figure}{Performance of traditional stylometric techniques average similarity in ads for vendors within and across Darknet datasets.
 \newline}
\label{fig:style_similarity}
\end{minipage}
}

The low similarity scores within and across datasets indicate the limited capabilities of traditional stylometric frameworks and suggest the need for mathematical models that can abstract features on higher levels. The low scores also serve as a sanity check indicating that vendors on Darknet use different vocabulary and styles in their ads within and across different markets, indicating the need for more profound feature-abstraction techniques. 

\subsubsection{Vendor Verification Task: Influence of advertisement frequency and trade categories on classifier's performance}

\begin{table}[h]
\begin{minipage}{\linewidth}
\centering
\resizebox{\columnwidth}{!}{%
    \begin{tabular}{|c|c|c|c|}
    \hline \textbf{Vendor} & \textbf{Ad Frequencies} & \textbf{Categories} & \textbf{F1-Score} \\
    \hline
    googleyed & 349 & 63 & 0.9340 \\ \hline
    etizolam & 186 & 59 & 0.9462 \\ \hline
    gotmilk & 842 & 48 & 0.9893 \\ \hline
    rinran & 437 & 47 & 0.9816 \\ \hline
    uhrwerk & 135 & 41 & 0.9925 \\ \hline
    citizen5 & 35 & 1 & 1.0000 \\ \hline
    corktech & 35 & 1 & 0.9714 \\ \hline
    sabinas & 26 & 1 & 0.9615 \\ \hline
    mrsupermario & 24 & 1 & 0.9615 \\ \hline
    emperium & 22 & 1 & 1.000 \\ \hline 
    \end{tabular}
    }
\caption{\label{tab:trade_categories} F1-score w.r.t vendor advertisement frequency and trade categories.}
\end{minipage}
\end{table}

\noindent The Alphabay-Dreams-Silk Road dataset consists of 272,696 unique ads and 3,896 vendors with 322 distinct categories. Table \ref{tab:trade_categories} illustrates the performance of our established BERT-cased classifier for five vendors selling trades across the most distinct categories and five vendors selling trades across only one category. As can be seen, the classifier's performance remains unaffected (more or less) with the number of trade categories and advertisement frequencies. The consistent performance suggests that despite the trade being conducted amongst different categories, Darknet vendors tend to advertise their products similarly, allowing our classifier to distinguish between unique writing styles from different vendors.

\subsubsection{Applying Knowledge Transfer: adapting to verify vendors from High Resource (HR) emerging markets} 
\label{app:hr_baseline}

\begin{table}[htp]
\centering
\resizebox{\columnwidth}{!}{%
\begin{tabular}{|l|l|l|l|}
\hline
\multicolumn{1}{|l|}{\textbf{Models}} & \multicolumn{1}{l|}{\textbf{Layer}} & \multicolumn{1}{l|}{\textbf{Micro-F1}} & \textbf{Macro-F1} \\ \hline
\multicolumn{4}{|c|}{\textit{Zero-Shot Baselines}} \\ \hline
\multicolumn{1}{|l|}{Architectural} & \multicolumn{1}{c|}{-} & \multicolumn{1}{l|}{0.7305} & 0.2173 \\ \hline
\multicolumn{1}{|l|}{Methodological} & \multicolumn{1}{c|}{-} & \multicolumn{1}{l|}{0.6498} & 0.1563 \\ \hline
\multicolumn{4}{|c|}{End-to-End Baselines} \\ \hline
\multicolumn{1}{|l|}{\textbf{BERT-cased}} & \multicolumn{1}{c|}{-} & \multicolumn{1}{l|}{\textbf{0.8750}} & \textbf{0.8700} \\ \hline
\multicolumn{1}{|l|}{BiGRU-Fasttext} & \multicolumn{1}{c|}{-} & \multicolumn{1}{l|}{0.6577} & 0.6539 \\ \hline
\multicolumn{4}{|c|}{\textit{Transfer Baselines}} \\ \hline
\multicolumn{1}{|l|}{\multirow{5}{*}{\textbf{\begin{tabular}[c]{@{}l@{}}Transfer-\\ BiGRU\end{tabular}}}} & \multicolumn{1}{l|}{Embedding} & \multicolumn{1}{l|}{0.6707} & 0.6698 \\ \cline{2-4} 
\multicolumn{1}{|l|}{} & \multicolumn{1}{l|}{Last} & \multicolumn{1}{l|}{0.7061} & 0.7153 \\ \cline{2-4} 
\multicolumn{1}{|l|}{} & \multicolumn{1}{l|}{Second-to-Last} & \multicolumn{1}{l|}{0.6992} & 0.6911 \\ \cline{2-4} 
\multicolumn{1}{|l|}{} & \multicolumn{1}{l|}{Weighted Sum All 12} & \multicolumn{1}{l|}{0.6698} & 0.6703 \\ \cline{2-4} 
\multicolumn{1}{|l|}{} & \multicolumn{1}{l|}{\textbf{Weighted Sum Last 4}} & \multicolumn{1}{l|}{\textbf{0.8065}} & \textbf{0.8177} \\ \hline
\end{tabular}%
}
\caption{Performance of Zero-Shot, End-to-End, and Transfer baselines on the Traderoute-Agora dataset.}
\label{tab:hr_baseline}
\resizebox{\columnwidth}{!}{%
\begin{tabular}{|c|c|c|c|}
\hline
\textbf{GPU} & \textbf{Models} & \multicolumn{1}{l|}{\textbf{\begin{tabular}[c]{@{}l@{}}Trainable\\ parameters\end{tabular}}} & \multicolumn{1}{l|}{\textbf{\begin{tabular}[c]{@{}l@{}}Training Time\\   (Hrs:Mins)\end{tabular}}} \\ \hline
\multirow{3}{*}{\begin{tabular}[c]{@{}c@{}}Tesla-\\ V100\\ (32 GB)\end{tabular}} & BERT-cased & 112M & 32:30 \\ \cline{2-4} 
 & BiGRU-Fasttext & 31M & 2:25 \\ \cline{2-4} 
 & Transfer-BiGRU & 42M & 17:23 \\ \hline
\end{tabular}%
}
\caption{Computational details of trained classifiers on the Traderoute-Agora dataset.}
\label{tab:hr_compuations}
\end{table}

\noindent In this research, we demonstrate the ability of our approach to adapt and verify migrating vendors from emerging LR markets using a compute-efficient network (transfer-BiGRU). Similar to the results presented in Section \ref{result:knowledge-transfer}, tables \ref{tab:hr_baseline} and \ref{tab:hr_compuations} demonstrate the performance and computational details of a transfer-BiGRU classifier on an HR emerging, Traderoute-Agora, dataset. As can be seen, despite the lesser trainable parameters and training time, our transfer-BiGRU underperforms compared to the end-to-end BERT-cased baseline. Therefore, we do not claim that our knowledge transfer approach scales to emerging vendors in HR Darknet markets.

\subsubsection{Seed Runs}
\noindent Due to limited resource constraints, we only analyze the effects of different initializations on our model's performance for the established benchmarks. As seen in Table \ref{tab:seed_runs}, the standard deviation, variance, and average performance suggest around 1\% influence of initialization on the model's performance. We report all the performance in this work based on our experiments conducted with a seed value of 1111.

\begin{table}[h]
\resizebox{\columnwidth}{!}{%
\begin{tabular}{|l|l|l|l|}
\hline
\multicolumn{1}{|c|}{\textbf{\begin{tabular}[c]{@{}c@{}}Seed\\ Value\end{tabular}}} & \multicolumn{1}{c|}{\textbf{\begin{tabular}[c]{@{}c@{}}BERT-cased\\ Alpha-Dreams-\\ Silk Dataset\end{tabular}}} & \multicolumn{1}{c|}{\textbf{\begin{tabular}[c]{@{}c@{}}BERT-cased\\ Valhalla-Berlusconi\\     Dataset\end{tabular}}} & \textbf{\begin{tabular}[c]{@{}l@{}}transfer-\\ BiGRU\end{tabular}} \\ \hline
\multicolumn{1}{|c|}{40} & \multicolumn{1}{c|}{0.8969} & \multicolumn{1}{c|}{0.8039} & \multicolumn{1}{c|}{0.7798} \\ \hline
100 & \multicolumn{1}{c|}{0.8824} & \multicolumn{1}{c|}{0.8278} & \multicolumn{1}{c|}{0.8005} \\ \hline
500 & \multicolumn{1}{c|}{0.8813} & \multicolumn{1}{c|}{0.7837} & \multicolumn{1}{c|}{17:23} \\ \hline
1100 & 0.8861 & 0.8089 & 0.8019 \\ \hline
\textbf{1111} & \textbf{0.9013} & \textbf{0.8290} & \textbf{0.8132} \\ \hline
Var. & 6.46 x 10\textasciicircum{}\{-5\} & 0.0002 & 0.0002 \\ \hline
Std. & 0.0080 & 0.0167 & 0.0143 \\ \hline
\textbf{Avg.} & \textbf{0.8896} & \textbf{0.8106} & \textbf{0.8035} \\ \hline
\end{tabular}%
}
\caption{Influence of different initialization on macro-F1 performance.}
\label{tab:seed_runs}
\end{table}

\subsubsection{Model Explanations} 
\label{app:explainability}

\begin{minipage}[h]{\columnwidth}
\includegraphics[width=\columnwidth]{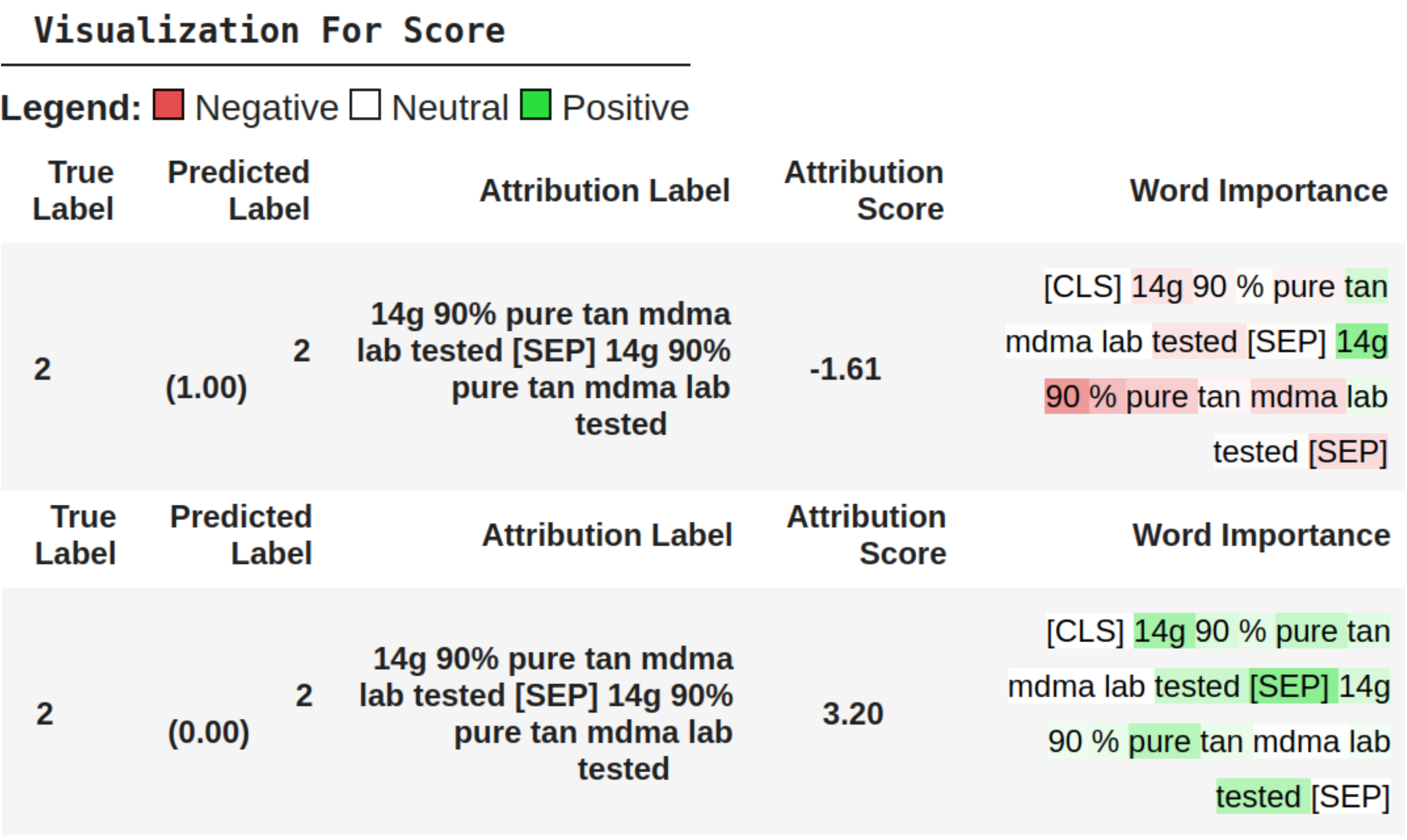}
\captionof{figure}{Inconsistency in model explanations within different explainability frameworks.\newline}
\label{fig:explainaibility_cross_platform}
\end{minipage}

\noindent We also conduct various word attributions-based explainability experiments on our BERT-cased methodological classifier to understand our model's decisions. Figure \ref{fig:explainaibility_cross_platform} illustrates the word attributions of the same advertisement from a vendor, "pckabml", generated through the \href{https://captum.ai/}{captum} \citep{kokhlikyan2020captum} and \href{https://pypi.org/project/transformers-interpret/}{transformers-interpret} \citep{Pierse_Transformers_Interpret_2021} frameworks. As can be seen, despite the ads being the same, different explainability frameworks generates different word attributions causing inconsistency in our explanations.

\noindent On the other hand, figure \ref{fig:explainaibility_across_vendors} illustrates the captum-based word attributions for similar ads from a vendor, "uridol". As can be seen, despite the similarity in ads and generating explanations from the same framework, we get different word attributions causing inconsistency in our explanations. We believe that computing the word attributions through the [CLS] token instead of the entire advertisement could be one of the reasons for these inconsistencies. While we do not clearly understand the reasoning behind the discrepancy in our explanations, we plan to investigate it in the future. \newline

\begin{minipage}{\columnwidth}
\includegraphics[ width=\columnwidth]{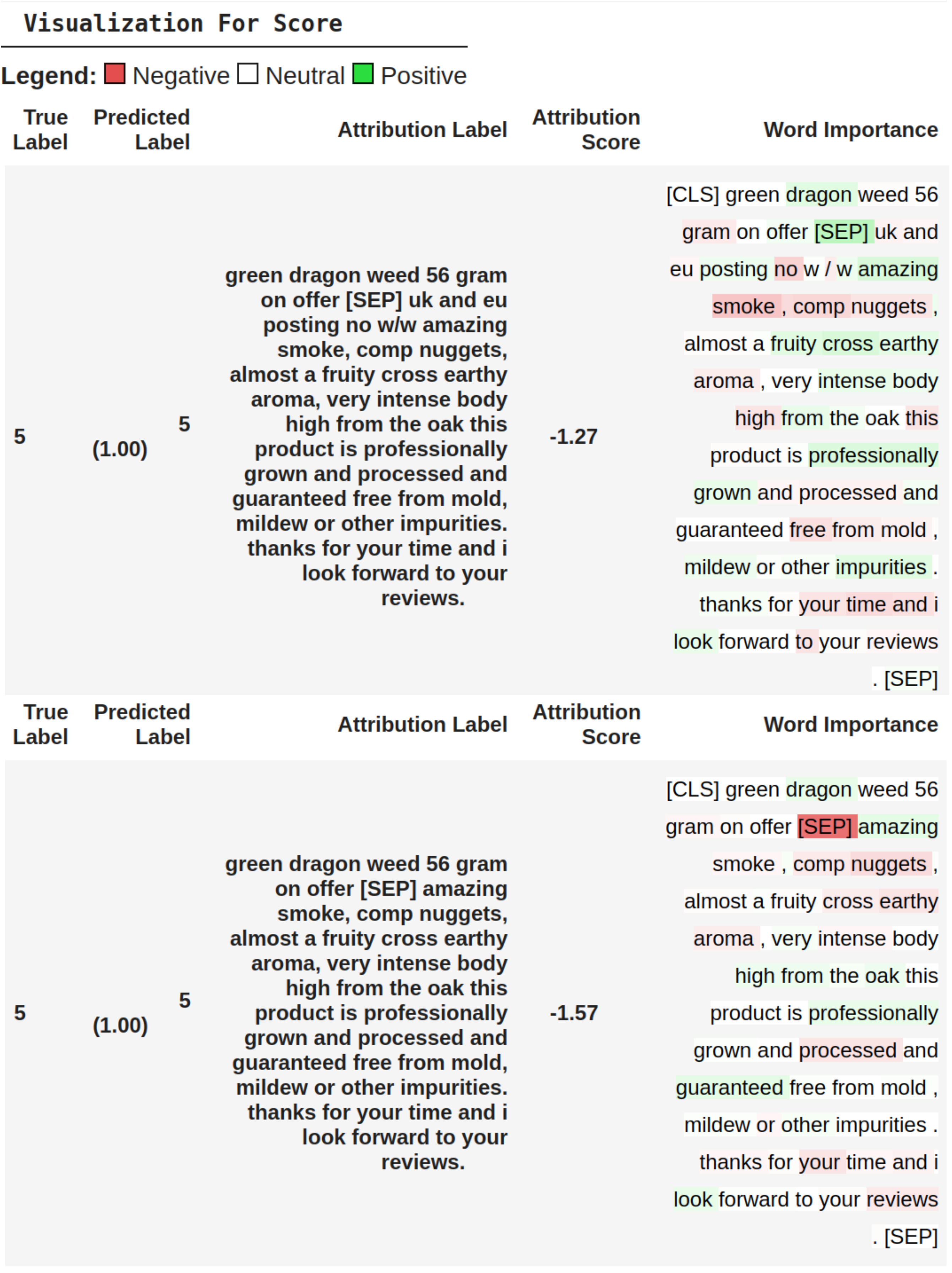}
\captionof{figure}{Inconsistency in model explanations for similar ads from the same vendor.\newline}
\label{fig:explainaibility_across_vendors}
\end{minipage}

\subsection{Infrastructure \& Schedule}
\label{appendix:infrastructure}
\paragraph{Data:} We perform our experiments using the standard splitting ratio of 0.75:0.05:0.20 ratio for the train, validation, and test dataset. 

\paragraph{Training:} We perform the training and evaluation of our Neural Networks on a single Tesla V100 GPU with 32 GBs of memory. The training and evaluation of statistical classifiers are performed on a server with one Intel Xeon Processor E5-2698 v4 and 512 GBs of RAM. Finally, we train our distilled transfer-BiGRU model for the Low-Resource setting on a GeForce-MX110 graphic card with 2 GBs of memory.

We use Adam optimizer with $\beta$1 = 0.9, $\beta$2 = 0.999, L2 weight decay of 0.01, and a learning rate of 0.001 with warm-up over the first 500 steps, and a linear decay.

\paragraph{Architectures \& Hyperparameters \footnote{All the models are implemented in python \cite{van1995python} using \href{https://scikit-learn.org/}{Sklearn} \cite{scikit-learn}, \href{https://pytorch.org/}{PyTorch} \cite{NEURIPS2019_9015}, and \href{https://huggingface.co/}{Hugging-face} \cite{wolf-etal-2020-transformers} frameworks.}:} We train all our statistical models using unigrams and bigrams features and balanced class weights. We experiment SVMs with both linear and Radial basis function (RBF) kernels, Random Forest with n\_estimators of 100 and 1000, max\_depth of 5, 10, and 20, and MLP with 100 layers and 100 neurons each. Finally, we evaluate our statistical models on the test dataset using a 5-fold nested cross-validation technique.

Our CNN architecture operates on sequences of n-grams characters extracted from the Darknet ads. We then pass the extracted embeddings through six convolutional with max-pooling and three fully connected layers. Inspired by \citep{zhang2016characterlevel}, we kept the input length to 1,014, dropout to 0.5 for the fully connected layers with 768 neurons each, a kernel size of 7 in the first two convolutional layers and 3 for the remaining layers. Finally, we set the filter size to 32 and train our models with a batch size of 32 until convergence.

The RNN architecture contains a two-layer Bidirectional-GRU model with two fully connected layers and fasttext embeddings. We first pack and pad the input sequence with variable length through a PyTorch function and then pass it to the embedding layer. After generating the text representation from the Bi-GRU layers, we finally pass the output through a softmax layer and perform classification over it. After some experimentations, we set the number of hidden units to 768, dropout to 0.65, batch size to 32, and trained the model until convergence.

Finally, we train several transformers models (BERT-base-cased, BERT-base-uncased, RoBERTa-base, and DistilBERT-base-cased) with a sequence classification head on top at a batch size of 32 \footnote{The maximum batch size allowed by our resources without running into memory issues.} for 40 epochs (due to computational reasons) for the architectural baselines and till convergence for the methodological baselines. We also train a BERT-base-uncased model on the language task for 20 epochs. All the transformer-based architectures are initialized from a pre-trained model checkpoint. 

\paragraph{Code Repository:} We encourage our readers to access and replicate our code through \href{https://anonymous.4open.science/r/VendorLink-9BE0/}{VendorLink}.

\paragraph{Computational Details:} Tables \ref{tab:training_time1} and \ref{tab:training_time2} presents details about the number of trainable parameters and execution time for all the trained models in the architectural and methodological baselines.

\begin{table}[H]
\centering
\resizebox{\columnwidth}{!}{%
\begin{tabular}{|c|c|c|}
\hline \textbf{Models (trained on} & \textbf{Trainable} & \textbf{Training} \\
\textbf{Dreams data)} & \textbf{parameters} & \textbf{time in hrs.} \\
\hline \hline
Multinomial &  & \\ 
Naive Bayes & - & 53:56 \\ \hline
Random Forest & - & 68:27 \\ \hline
Logistic Regression & - & 79:42 \\ \hline
SVM & - & 81:08 \\ \hline
MLP & - & 94:18 \\ \hline \hline
Character-CNN & 16M & 0:54 \\ \hline
GRU-Fasttext & 39M & 1:12 \\ \hline \hline
BERT & 110M & 25:14 \\ \hline
RoBERTa & 125M & 23:40 \\ \hline
DistilBERT & 68M & 17:57 \\ \hline
\end{tabular}
}
\caption{\label{tab:training_time1} Number of trainable parameters and training time for architectural baselines.}
\end{table}

\begin{table}[H]
\centering
\resizebox{\columnwidth}{!}{%
\begin{tabular}{|c|c|c|}
\hline \textbf{Models (trained on} & \textbf{Trainable} & \textbf{Training} \\
\textbf{Alphabay-Dreams} & \textbf{parameters} & \textbf{time in hrs.} \\
\textbf{-Silk Road dataset)} & \textbf{} & \textbf{} \\ \hline \hline
BERT-uncased & 111M & 67:02 \\ \hline
BERT-cased & 112M & 66:58 \\ \hline \hline
DarkBERT-LM & 108M & 156:14 \\ \hline
DarkBERT Classifier & 112M  & 49:39 \\ \hline
Adapter BERT & 4M & 51:00 \\ \hline
\end{tabular}
}
\caption{\label{tab:training_time2} Number of trainable parameters and training time for methodological baselines.}
\end{table}

\paragraph{Evaluation Metrics:} We evaluate our trained classifiers against accuracy, micro-average F1, and macro-average F1 (commonly known as macro-F1 and micro-F1) using the classification report from scikit-learn. We argue that macro-F1 computes the score independently for each class and then takes the average (treating majority and minority classes equally). Given the class imbalance we have in our dataset, we heavily emphasize our trained models' performance on macro-F1 scores. Furthermore, we evaluate the BERT-base language model on loss and perplexity. Finally, we use Centered Kernel Alignment (CKA) to evaluate and compute correspondences between our methodological baseline representations before and after finetuning. 

\subsection{Assumptions}
\label{app:assumptions}
This work applies a lower-case transformation to the vendor names during the pre-processing step and assumes vendor accounts "agentq" and "AgentQ" to be from the same entity. However, in reality, these entities can refer to two different vendors. Additionally, we train our classifier in a multi-class classification setting, assuming that ads correspond to only one individual vendor account. However, our experiments uncover the existence of copycats on Darknet markets. In reality, it is always possible for multiple vendors to co-exist with similar vendor names; hence, any supervised approach will only generate skew results. In the future, we plan to look toward contrastive learning approaches \citep{https://doi.org/10.48550/arxiv.2107.10137, Zhou2021ContrastiveOD, wegmann-etal-2022-author} to avoid these assumptions. 

\subsection{Limitations}
\label{app:limitations}
\paragraph{Architectural limitations:} This research establishes a BERT-base-cased classifier to verify migrating vendors across existing and emerging Darknet markets. While we acknowledge that using a bigger BERT model with a sliding window may improve our classification's performance, given the resources at our disposal, we decided against it. Moreover, as mentioned earlier, most of the ads used in this research are in English, with a few exceptions where the vendors use multiple languages. Therefore, applying a multilingual transformer-based model to the classification task \citep{Wang2021PracticalTM} can improve our approach's performance.

\paragraph{Unsupervised and HR settings:} As described in the appendix section \ref{app:assumptions}, the core of our approach lies in the availability of gold labels. VendorLink utilizes the supervised pre-training step to perform knowledge transfer and text-similarity tasks. Therefore, our approach suffers a significant limitation in the absence of these ground labels / unsupervised settings. Furthermore, as described in \ref{app:hr_baseline}, our approach could not scale well to verify vendor migrants in HR emerging datasets. In the future, we plan to expose VendorLink to contrastive learning approaches to learn universal representations and overcome the problem. 

\paragraph{Diverse Advertisements:} In the semi-supervised task, we compute the likelihood of two vendor accounts being from the same entity by calculating the similarity between the advertisements of the two vendors. Since one of the novelties of this research lies in the direction of End-to-End training, we have avoided using handcrafted labels for applying content control to generate content-independent style representation. However, as explained in section \ref{exp:representational_similarity}, an advertisement from the drug category can be very different from that of the weapon category. Therefore, in the future, we plan to train another classifier to classify Darknet advertisements into different trade categories before performing the vendor-verification task. 

\paragraph{XAI limitations:} eXplainaible Artificial Intelligence (XAI) is integral in promoting trust and understanding amongst the end-users. From LEA's perspective, its absence can be viewed as arguably negligent and unreliable. While we acknowledge that our approach currently lacks an XAI feature, in future, we plan to build upon our experiments in \ref{app:explainability} and establish a reliable approach for understanding and explaining our model's decision. \newpage

  
  
        
  

\end{document}